# Impact of squark generation mixing on the search for squarks and gluinos at LHC

**Keisho Hidaka**[1]

*Department of Physics, Tokyo Gakugei University, Koganei, Tokyo 184-8501, Japan*
*E-mail:* `hidaka@u-gakugei.ac.jp`

**Alfred Bartl, Karl Hohenwarter-Sodek, Thomas Kernreiter**

*Faculty of Physics, University of Wien, A-1090 Vienna, Austria*

**Helmut Eberl, Walter Majerotto**

*IHEP der Österreichischen Akademie der Wissenschaften, A-1050 Vienna, Austria*

**Björn Herrmann**

*Deutsches Elektronen-Synchrotron (DESY), Theory Group, D-22603 Hamburg, Germany*

**Werner Porod**

*Institut für Theoretische Physik und Astrophysik, Universität Würzburg, D-97074 Würzburg, Germany*

We study the effect of squark-generation mixing on production and decays of squarks and gluinos at LHC in the Minimal Supersymmetric Standard Model (MSSM). We show that the mixing effects can be very large in a significant range of the squark-generation mixing parameters despite the very strong constraints on quark-flavour violation (QFV) from experimental data on B mesons. We find that under favourable conditions the QFV decay branching ratio $B(\tilde{g} \to c\bar{t}(t\bar{c})\tilde{\chi}_1^0)$ can be as large as about 50%, which may lead to significant QFV signals at LHC. We also find that the squark generation mixing can result in a novel multiple-edge (3- or 4-edge) structure in the charm-top quark invariant mass distribution. Further we show that the two lightest up-type squarks $\tilde{u}_{1,2}$ can have very large branching ratios for the decays $\tilde{u}_i \to c\tilde{\chi}_1^0$ and $\tilde{u}_i \to t\tilde{\chi}_1^0$ simultaneously due to the mixing effect, resulting in QFV signals '$pp \to c\bar{t}(t\bar{c})$ + missing-$E_T$ + X' at a significant rate at LHC. These remarkable signatures could provide a powerful test of supersymmetric QFV at LHC and could have an important impact on the search for squarks and gluinos and the determination of the MSSM parameters at LHC.

---

[1] Speaker

## 1. Introduction

The decays of gluinos and squarks are usually assumed to be quark-flavour conserving. However, the squarks are not necessarily quark-flavour eigenstates and they are in general mixed. In this case quark-flavour violating (QFV) decays of squarks and gluinos could occur. The effect of QFV in the squark sector on reactions at colliders has been studied only in a few publications. In this article based on [1, 2] we study the effect of QFV due to the mixing of charm-squark and top-squark on production and decays of squarks and gluinos at LHC in the general Minimal Supersymmetric Standard Model (MSSM).

## 2. Squark mixing with flavour violation and constraints

The most general up-type squark mass matrix including left-right mixing as well as quark-flavour mixing in the super-CKM basis of $\tilde{u}_{0\gamma} = (\tilde{u}_L, \tilde{c}_L, \tilde{t}_L, \tilde{u}_R, \tilde{c}_R, \tilde{t}_R)$, $\gamma = 1,\ldots,6$ is [1]:

$$M_{\tilde{u}}^2 = \begin{pmatrix} M_{\tilde{u}LL}^2 & M_{\tilde{u}LR}^2 \\ M_{\tilde{u}RL}^2 & M_{\tilde{u}RR}^2 \end{pmatrix}, \tag{1}$$

$$\left(M_{\tilde{u}LL}^2\right)_{\alpha\beta} = M_{Q_u\alpha\beta}^2 + \left[(\tfrac{1}{2} - \tfrac{2}{3}\sin^2\theta_W)\cos 2\beta\, m_Z^2 + m_{u_\alpha}^2\right]\delta_{\alpha\beta}, \tag{2}$$

$$\left(M_{\tilde{u}RR}^2\right)_{\alpha\beta} = M_{U\alpha\beta}^2 + \left[\tfrac{2}{3}\sin^2\theta_W \cos 2\beta\, m_Z^2 + m_{u_\alpha}^2\right]\delta_{\alpha\beta}, \tag{3}$$

$$\left(M_{\tilde{u}RL}^2\right)_{\alpha\beta} = (v_2/\sqrt{2})A_{U\beta\alpha} - m_{u_\alpha}\mu^*\cot\beta\,\delta_{\alpha\beta}. \tag{4}$$

$M_{\tilde{u}LR}^2$ is the hermitean conjugate of $M_{\tilde{u}RL}^2$. $M_{Q_u}^2$ and $M_U^2$ are the hermitean soft-supersymmetry (SUSY)-breaking mass matrices for the left and right up-type squarks, respectively. Note that $M_{Q_u}^2 = K \cdot M_Q^2 \cdot K^{-1}$, where $M_Q^2$ is the soft-SUSY-breaking mass matrix for the left down-type squarks and $K(\approx 1)$ is the CKM matrix. $m_{u_\alpha}$ ($u_\alpha = u, c, t$) are the physical quark masses. The mass eigenstates $\tilde{u}_i$, $i = 1,\ldots,6$, are given by $\tilde{u}_i = R_{i\alpha}^{\tilde{u}}\tilde{u}_{0\alpha}$. The mixing matrix $R^{\tilde{u}}$ is obtained by a unitary transformation $R^{\tilde{u}}M_{\tilde{u}}^2 R^{\tilde{u}-1} = diag(m_{\tilde{u}_1},\ldots,m_{\tilde{u}_6})$, where $m_{\tilde{u}_i} < m_{\tilde{u}_j}$ for $i < j$. We define the QFV parameters $\delta_{\alpha\beta}^{uLL}$ and $\delta_{\alpha\beta}^{uRR}$ ($\alpha \neq \beta$) as follows: $\delta_{\alpha\beta}^{uLL} \equiv M_{Q\alpha\beta}^2/\sqrt{M_{Q\alpha\alpha}^2 M_{Q\beta\beta}^2}$, $\delta_{\alpha\beta}^{uRR} \equiv M_{U\alpha\beta}^2/\sqrt{M_{U\alpha\alpha}^2 M_{U\beta\beta}^2}$. The down-type squark mass matrix can be analogously parametrized as the up-type squark mass matrix.

We impose the following conditions on the MSSM parameter space in order to respect experimental and theoretical constraints which are described in detail in [2]:
    (i) Constraints from the B-physics experiments [3], such as
        $2.92 \cdot 10^{-4} < B(b \to s\gamma) < 4.22 \cdot 10^{-4}$, $|\Delta M_{B_s}^{SUSY} - 17.77| < 3.31\, ps^{-1}$ and those from
        further rare B-decays.
    (ii) The experimental limit on SUSY contributions to the electroweak $\rho$ parameter [3].
    (iii) The LEP and Tevatron limits on the SUSY particle masses [3].

(iv) Vacuum stability conditions for the trilinear couplings $A_{U\alpha\beta}$ and $A_{D\alpha\beta}$ [3]. Conditions (i) and (iv) strongly constrain the 2nd and 3rd generation squark mixing parameters $M^2_{Q23}$, $M^2_{D23}$, $A_{U23}$, $A_{D23}$ and $A_{D32}$.

## 3. Quark flavour violating gluino decays

Possible two-body decay modes of the squarks and gluino in our study are:
$$\tilde{g} \to \tilde{u}_i u_k,\ \tilde{d}_i d_k;\quad \tilde{u}_i \to u_k \tilde{\chi}^0_n,\ d_k \tilde{\chi}^+_m,\ \tilde{d}_j W^+,\ \tilde{u}_j Z^0,\ \tilde{u}_j h^0,$$
where $\tilde{\chi}^0_n, \tilde{\chi}^+_m$ and $h^0$ are neutralinos, charginos and the lightest Higgs boson, respectively. We take $\tan\beta, m_{A^0}, M_{1,2}, m_{\tilde{g}}, \mu, M^2_{Q\alpha\beta}, M^2_{U\alpha\beta}, M^2_{D\alpha\beta}, A_{U\alpha\beta}$ and $A_{D\alpha\beta}$ as the basic MSSM parameters at the weak scale, where $m_{A^0}$ is the CP-odd Higgs mass, and $M_2$ and $M_1$ are the SU(2) and U(1) gaugino masses, respectively. We take the following scenario as a reference scenario with QFV within the reach of LHC (All mass parameters are in GeV.):

$\tan\beta$ =10, $m_{A^0}$ =800, $M_1$ =139, $M_2$ =264, $m_{\tilde{g}}$ =800, $\mu$ =1000, $M^2_{Q11}$ = $(920)^2$, $M^2_{Q22}$ = $(880)^2$, $M^2_{Q33}$ = $(840)^2$, $M^2_{Q12}$ = $M^2_{Q13}$ = 0, $M^2_{Q23}$ = $(224)^2$, $M^2_{U11}$ = $(820)^2$, $M^2_{U22}$ = $(600)^2$, $M^2_{U33}$ = $(580)^2$, $M^2_{U12}$ = $M^2_{U13}$ = 0, $M^2_{U23}$ = $(224)^2$, $M^2_{D11}$ = $(830)^2$, $M^2_{D22}$=$(820)^2$, $M^2_{D33}$=$(810)^2$, $M^2_{D12}$=$M^2_{D13}$=$M^2_{D23}$= 0, and all of $A_{U\alpha\beta}$ and $A_{D\alpha\beta}$ are set to zero. In this scenario satisfying all the conditions (i)-(iv) we have (Masses are in GeV.):

$$m_{\tilde{u}_1}=558,\ m_{\tilde{u}_2}=642,\ \tilde{u}_1 \cong 0.728\tilde{c}_R - 0.685\tilde{t}_R,\ \tilde{u}_2 \cong -0.686\tilde{c}_R - 0.727\tilde{t}_R,\ m_{\tilde{\chi}^0_1}=138, \quad (5)$$

$$B(\tilde{g} \to c t \tilde{\chi}^0_1) = \sum_{i=1,2}[B(\tilde{g} \to \tilde{u}_i c)B(\tilde{u}_i \to t\tilde{\chi}^0_1) + B(\tilde{g} \to \tilde{u}_i t)B(\tilde{u}_i \to c\tilde{\chi}^0_1)] = 0.46, \quad (6)$$

$$B(\tilde{u}_1 \to c\tilde{\chi}^0_1) = 0.58,\ B(\tilde{u}_1 \to t\tilde{\chi}^0_1) = 0.40,\ B(\tilde{u}_2 \to c\tilde{\chi}^0_1) = 0.50,\ B(\tilde{u}_2 \to t\tilde{\chi}^0_1) = 0.47. \quad (7)$$

Here $B(\tilde{g} \to ct\tilde{\chi}^0_1) \equiv B(\tilde{g} \to c\bar{t}\tilde{\chi}^0_1) + B(\tilde{g} \to \bar{c}t\tilde{\chi}^0_1)$. The reason for the very large QFV gluino decay branching ratio in Eq.(6) is as follows: The gluino decays into squarks other than $\tilde{u}_{1,2}$ are kinematically forbidden, and $\tilde{u}_1$, $\tilde{u}_2$ are strong mixtures of $\tilde{c}_R$ and $\tilde{t}_R$ due to the large $\tilde{c}_R$-$\tilde{t}_R$ mixing term $M^2_{U23}$ (= $(224 \text{ GeV})^2$) in this scenario. Here note that $\tilde{u}_{1,2}(\sim \tilde{c}_R + \tilde{t}_R)$ couple to $\tilde{\chi}^0_1 (\cong \tilde{B}^0)$ and practically do not couple to $\tilde{\chi}^0_2 (\cong \tilde{W}^0)$, $\tilde{\chi}^\pm_1 (\cong \tilde{W}^\pm)$, and that $\tilde{\chi}^0_{3,4}$, $\tilde{\chi}^\pm_2$ are very heavy in this scenario. Here $\tilde{B}^0$ and $\tilde{W}^{0,\pm}$ are the U(1) and SU(2) gauginos, respectively.

In Fig.1 we show contours of $B(\tilde{g} \to ct\tilde{\chi}^0_1)$ in the $\delta^{uLL}_{23}$-$\delta^{uRR}_{23}$ plane where all of the conditions (i)-(iv) except the $b \to s\gamma$ constraint are satisfied. All basic parameters other than $M^2_{Q23}$ and $M^2_{U23}$ are fixed as in our reference scenario specified above. We see that $B(\tilde{g} \to ct\tilde{\chi}^0_1)$ increases quickly with increase of the $\tilde{c}_R$-$\tilde{t}_R$ mixing parameter $|\delta^{uRR}_{23}|$ and can be very large (up to ~ 50 %) in a significant part of the $\delta^{uLL}_{23}$-$\delta^{uRR}_{23}$ plane allowed by all of the conditions (i)-(iv) including the $b \to s\gamma$ constraint. From Fig.1 we find that the possibility of the large QFV effect can not be excluded by the $b \to s\gamma$ constraint even if the experimental error of $B(b \to s\gamma)$ becomes very small. The other QFV parameter dependences of $B(\tilde{g} \to ct\tilde{\chi}^0_1)$ are also shown in [1].

The signature of the decay $\tilde{g} \to ct\tilde{\chi}_1^0$ would be '(charm-) jet + top-quark + missing-energy'. We have also studied the effect of the squark generation mixing on the invariant mass distributions of the two quarks from the gluino decay at LHC. We have found that it can result in novel and characteristic edge structures in the distributions. In particular, multiple-edge (3- or 4-edge) structures can appear in the distribution of the invariant mass of the charm-top quark system. These edge structures are due to the fact that the gluino decays occur via an intermediate real squark.

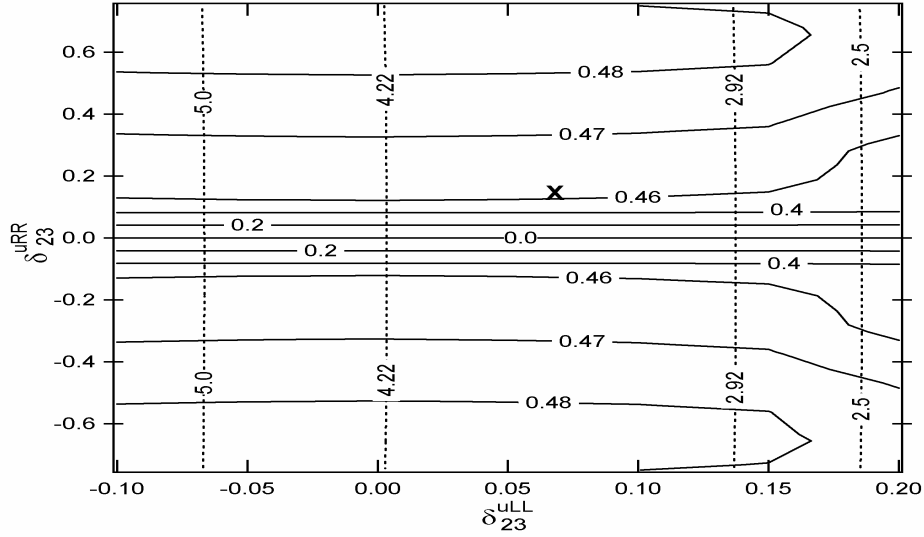

**Figure 1**: Contours of the QFV decay branching ratio $B(\tilde{g} \to ct\tilde{\chi}_1^0)$ (solid lines) in the $\delta_{23}^{uLL}$ - $\delta_{23}^{uRR}$ plane where all of the conditions (i)-(iv) except the $b \to s\gamma$ constraint are satisfied. Contours of $10^4 \cdot B(b \to s\gamma)$ (dashed lines) are also shown. The condition (i) requires $2.92 < 10^4 \cdot B(b \to s\gamma) < 4.22$. The point "x" of ($\delta_{23}^{uLL}, \delta_{23}^{uRR}$) = (0.068, 0.144) corresponds to our reference scenario.

## 4. Quark flavour violating squark decays

The large $\tilde{c}_R$ - $\tilde{t}_R$ mixing can also lead to large $B(\tilde{u}_i \to c\tilde{\chi}_1^0)$ and $B(\tilde{u}_i \to t\tilde{\chi}_1^0)$ ($i = 1,2$) (see Eq.(7)), which may result in a sizable rate for the following QFV signals at LHC [2]:

$$pp \to \tilde{u}_{1,2}\overline{\tilde{u}}_{1,2} X \to c\bar{t}\tilde{\chi}_1^0\tilde{\chi}_1^0 X, t\bar{c}\tilde{\chi}_1^0\tilde{\chi}_1^0 X, \qquad (8)$$

where X contains only beam jets and the $\tilde{\chi}_1^0$'s give rise to missing transverse energy $E_T^{mis}$. The $\delta_{23}^{uRR}$ dependence of the corresponding cross sections given by

$$\sigma_{ct}^{ii} \equiv \sigma(pp \to \tilde{u}_i\overline{\tilde{u}}_i X \to c\bar{t}\tilde{\chi}_1^0\tilde{\chi}_1^0 X) + \sigma(pp \to \tilde{u}_i\overline{\tilde{u}}_i X \to t\bar{c}\tilde{\chi}_1^0\tilde{\chi}_1^0 X) \qquad (9)$$

is shown for our reference scenario in Fig.2. We see that the QFV cross sections can be quite sizable in a wide allowed range of $\delta_{23}^{uRR}$. For an integrated luminosity of $L$=100 fb$^{-1}$ at LHC with 14 TeV one would expect more than $10^4$ events of the QFV signals.

The signature of the QFV processes would be '(charm-) jet + (anti) top-quark + $E_T^{mis}$ + X', where X contains only beam jets. The most important SUSY background would be due to the QFC production $pp \to \tilde{u}_i\overline{\tilde{u}}_i X \to t\bar{t}\tilde{\chi}_1^0\tilde{\chi}_1^0 X$, where one W-boson stemming from a top-quark decays hadronically and the other one decays leptonically with the charged lepton being missed or mis-identified. The most important SM background would be top-quark pair production $pp \to t\bar{t}Z^0 X \to t\bar{t}\nu\bar{\nu}X$, where one of the W-bosons from the top-quarks decays leptonically with the charged lepton being not detected. Single top-quark production $pp \to "W^+" Z^0 X \to t\bar{b}\nu\bar{\nu}X$ also could be a SM background. However the cross sections of these SM background processes would be very small because they involve weak processes. Detailed Monte Carlo studies including background processes and detector effects are necessary to identify the parameter region where the proposed QFV signal is observable with sufficient significance.

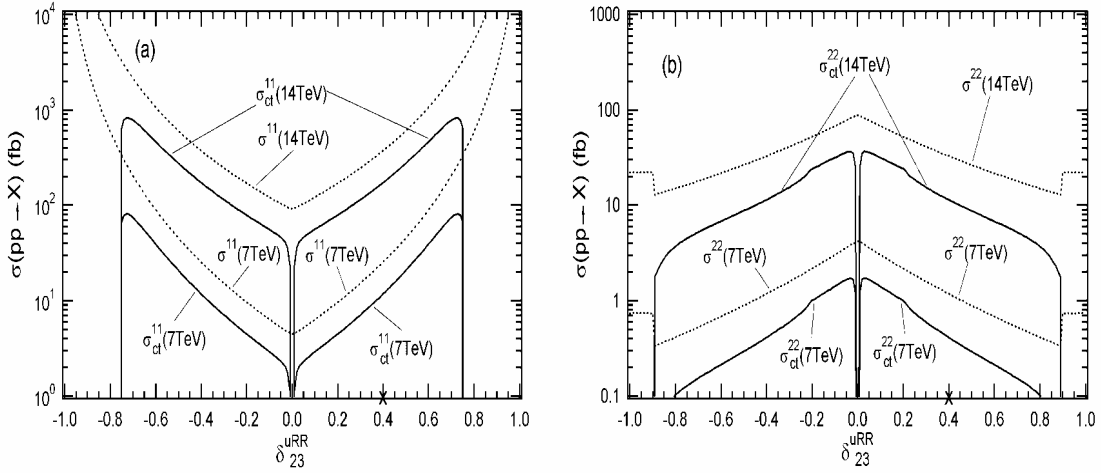

**Figure 2**: $\delta_{23}^{uRR}$ dependences of (a) $\sigma^{11} \equiv \sigma(pp \to \tilde{u}_1 \bar{\tilde{u}}_1 X)$, $\sigma_{ct}^{11}$ and (b) $\sigma^{22} \equiv \sigma(pp \to \tilde{u}_2 \bar{\tilde{u}}_2 X)$, $\sigma_{ct}^{22}$ at $E_{cm}$ = 7 and 14 TeV. All basic parameters other than $M_{U23}^2$ are fixed as in our reference scenario specified in the text. The shown range of $\delta_{23}^{uRR}$ is the whole range allowed by the conditions (i) to (iv) given in the text.

We remark that for the QFV scenarios based on the mSUGRA scenarios such as SPS1a' we have obtained results similar to those presented in this article, as shown in [1, 2].

## 5. Conclusion

Our analyses shown here suggest that one should take into account the possibility of significant contributions from QFV production and decays in the squark and gluino search at LHC. One should also include the QFV squark parameters (i.e. the squark generation mixing parameters) in the basic MSSM parameter determination at LHC.